\documentclass[aps,prb,twocolumn,floatfix]{revtex4}
\usepackage{graphicx}
\usepackage{nature}

\begin{document}

\title{Tailoring ferromagnetic chalcopyrites}
\author{Steven C. Erwin and Igor \v{Z}uti\'c}
\affiliation{Center for Computational Materials Science, Naval Research Laboratory,
Washington, D.C. 20375}
\date{\today}

\maketitle 

{\bf If magnetic semiconductors are ever to find wide application in real
spintronic devices, their magnetic and electronic properties will
require tailoring in much the same way that band gaps are engineered
in conventional semiconductors.  Unfortunately, no systematic understanding
yet exists of how, or even whether, properties such as Curie
temperatures and band gaps are related in magnetic
semiconductors. Here we explore theoretically these and other
relationships within 64 members of a single materials class, the
Mn-doped II-IV-V$_2$ chalcopyrites, three of which are already known
experimentally to be ferromagnetic semiconductors. Our
first-principles results reveal a variation of magnetic properties
across different materials that cannot be explained by either of the
two dominant models of ferromagnetism in semiconductors. Based on our
results for structural, electronic, and magnetic properties, we
identify a small number of new stable chalcopyrites with excellent prospects
for ferromagnetism.}

One of the tantalizing promises of dilute magnetic semiconductors is
to conjoin, in a single material, the advantages of nonvolatility and
bandgap
engineering\cite{ohno_science_1998a,maekawa2002a,zutic_rev_mod_phys_2004a}.
In normal (nonmagnetic) semiconductors, precise tailoring of band
structure is possible due to our detailed understanding---both
empirical and theoretical---of the mechanisms underlying band
formation.  Unlike the situation for normal semiconductors, however,
there does not yet exist any comprehensive understanding of how the
magnetic properties of magnetic semiconductors are related to their
structural and electronic properties.

For example, there is no firm understanding of the relationship
between the properties of the host semiconductor and the Curie
temperatures attainable by doping with magnetic impurities such as Mn.
Experimentally, a number of materials issues---including Mn content,
compensation, and phase purity---are not yet sufficiently under
control to permit a systematic description across different
semiconductors. Theoretically, there is an emerging consensus that
while the ferromagnetism originates from the interaction between
itinerant electrons (or holes) and localized electrons, there is no
single model that can systematically relate the physical and
electronic properties of the host semiconductor to its resulting
magnetic properties when doped\cite{dietl_nature_mater_2003a}.  Models
based on the weak- and strong-coupling limits of this interaction (the
Zener\cite{dietl_science_2000a,konig_phys_rev_lett_2000a} and
double-exchange\cite{chattopadhyay_phys_rev_lett_2001a,nagaev2002a} models,
respectively) do make specific predictions, but it is increasingly
evident that real materials are not sufficiently well described by
either limit for these predictions to be reliable across different
hosts.    
    
\begin{figure}[b]
\resizebox{86mm}{!}{\includegraphics{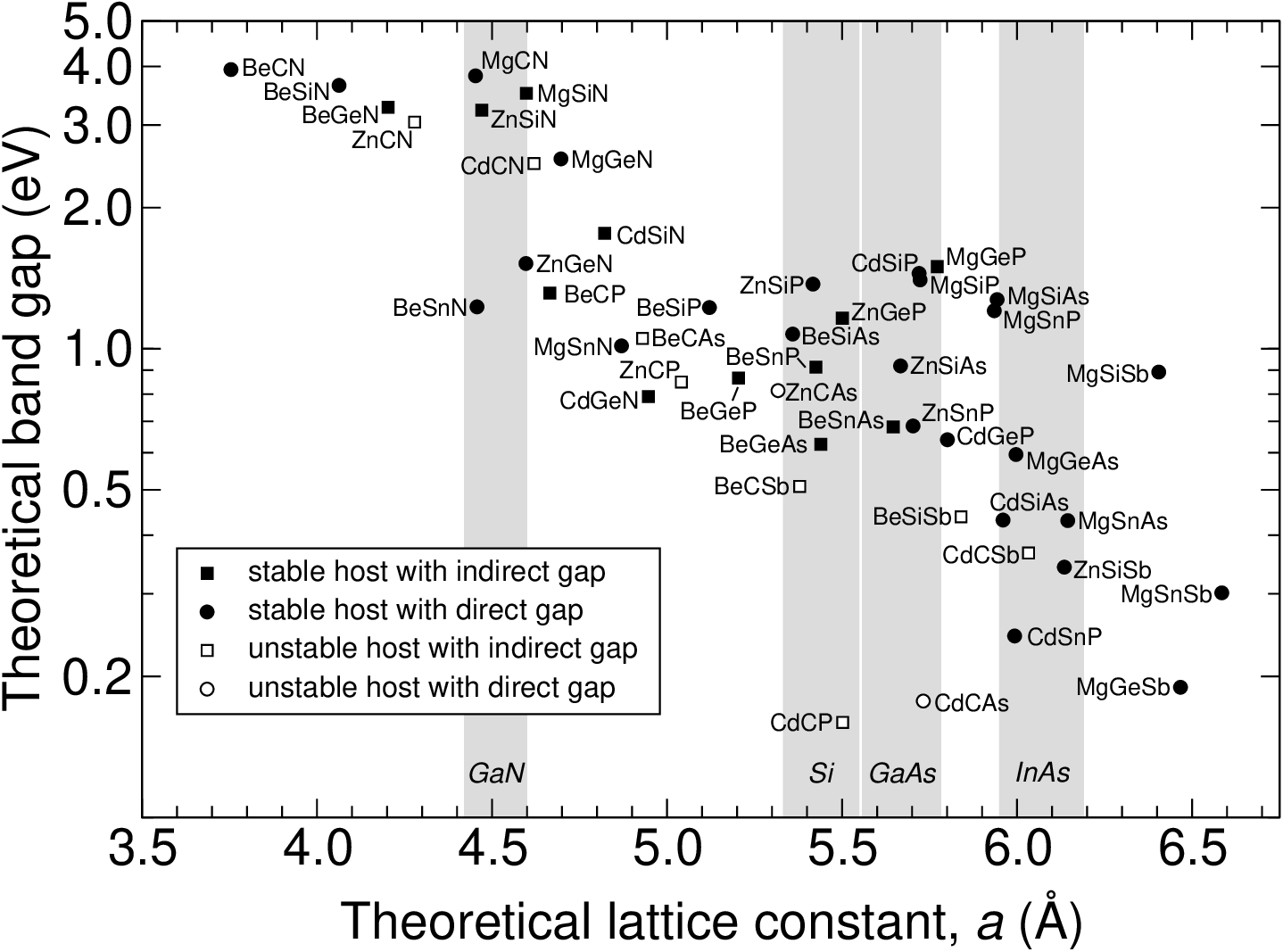}}
\caption{Theoretical band gap vs.~lattice constant for the
II-IV-V$_2$ chalcopyrites considered here.  Shaded areas indicate
hosts that are expected to be closely ($\pm 2\%$) lattice matched to
either GaN, Si, GaAs, or InAs. Filled symbols denote stable host
compounds (negative enthalpies of formation). Chalcopyrites whose
calculated gaps are zero (see text) are not shown.}
\end{figure}

Here we address this problem by exploring theoretically these
relationships within a single materials class: the II-IV-V$_2$
chalcopyrites consisting of the 64 possible combinations of
II=(Be,Mg,Zn,Cd), IV=(C,Si,Ge,Sn), and V=(N,P,As,Sb).  Three of
these---CdGeP$_2$, ZnGeP$_2$, and ZnSnAs$_2$---were recently shown to
become ferromagnetic upon Mn doping, with remarkably high Curie
temperatures\cite{medvedkin_japanese_journal_of_applied_physics_2000a,medvedkin_journal_of_crystal_growth_2002a,cho_phys_rev_lett_2002a,choi_solid_state_comm_2002a}.
It is an open question whether other chalcopyrites can be made
ferromagnetic; nor is it known which properties of the host
chalcopyrite are important for ferromagnetism and, in particular, for
high Curie temperatures.  These are the central issues we address in
this work.
 
For each chalcopyrite, we use density-functional theory (DFT) in the
generalized-gradient
approximation\cite{perdew_phys_rev_b_1992a,kresse_phys_rev_b_1993a,kresse_phys_rev_b_1996a}
to compute several properties of the undoped hosts: equilibrium
crystal structure, i.e.~lattice parameters $a$ and $c$ and the single
internal coordinate; phase stability, as determined by the enthalpy of
formation; and electronic band gap, which may be direct or indirect.
A comprehensive survey of the magnetic properties within such a large
class of materials is impractical, so we consider here only a few key
quantities that control magnetism. Since Mn is isoelectronic with the
group-II elements, it is believed that carrier-mediated ferromagnetism
in the II-IV-V$_2$ chalcopyrites is favored when Mn occupies the
group-IV site, where it is likely to be electrically
active\cite{mahadevan_phys_rev_lett_2002a}. For this reason we first
explore the competition between substitutional Mn doping on the II and
IV sites by computing the Mn impurity formation energy on each site.
Second, we address whether occupation of the group-IV sites is indeed
likely to lead to a ferromagnetic phase; in particular, for each host
we compute the Heisenberg spin coupling for an experimentally
reasonable Mn concentration. The magnitude (and sign) of this coupling
should serve as a useful roadmap for the experimental exploration of
new ferromagnetic chalcopyrites.

Of the 64 chalcopyrites considered here, about 1/3 are known to exist
and have been studied experimentally.  A few crystallize in both the
chalcopyrite and sphalerite structure (e.g.~CdGeP$_2$ and ZnSnSb$_2$),
while two take the wurzite structure (BeSiN$_2$ and ZnGeN$_2$); for
simplicity we treat all 64 in the chalcopyrite structure. For those
with measured lattice parameters, we can compare our theoretical
predictions. As expected from DFT calculations, $a$ is given very
accurately (within $\sim$1\%), while $c$ is less so ($\lesssim$15\%).  As
Fig.~1 shows, nearly half of the 64 chalcopyrites enjoy a close
lattice match to an important mainstream semiconductor.

As with all semiconductors, DFT in the generalized-gradient
approximation significantly underestimates band gaps.  For
chalcopyrites whose measured gaps are in the range 1--2 eV, the DFT
values are approximately 1 eV smaller than measured, consistent with
previous findings\cite{continenza_phys_rev_b_1992a}. Hence the theoretical
gaps shown in Fig.~1 must be viewed with caution: the true gaps are
likely to be at least 1 eV larger than predicted.  Likewise, hosts
that have no gap within DFT may in reality have a gap (which is likely
to be 1 eV or smaller). These caveats aside, the trend in
Fig.~1 is standard for covalently bonded semiconductors: smaller
lattice constants are associated with larger band gaps.  In
particular, the nitrides have the smallest lattice constants and
largest gaps, while the antimonides have the largest lattice constants
and smallest gaps.  From the perspective of our survey, the range and
distribution of lattice constants and band gaps is satisfyingly large,
guaranteeing our roadmap good coverage of this unexplored territory.

\begin{figure}[b]
\resizebox{86mm}{!}{\includegraphics{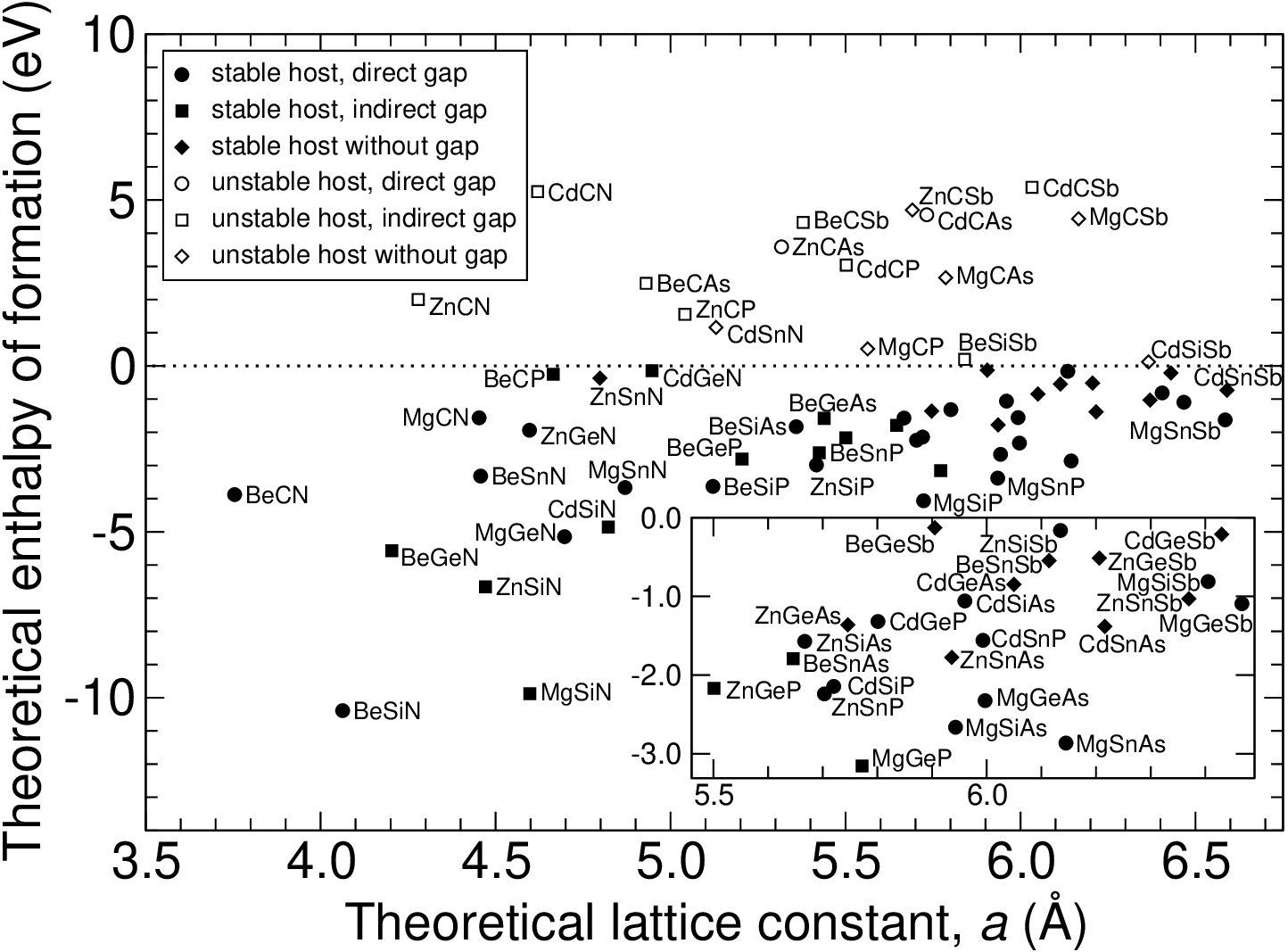}}
\caption{Theoretical enthalpy of formation vs.~lattice constant for
II-IV-V$_2$ chalcopyrites.  Negative enthalpy implies a stable host compound.}
\end{figure} 

Not all of the chalcopyrites considered here are thermodynamically
stable, even in the absence of Mn. Fig.~2 shows the calculated
enthalpies of formation, 
$\Delta H_f(\textrm{II-IV-V}_2)=E(\textrm{II-IV-V}_2)-E(\textrm{II})-E(\textrm{IV})-2E(\textrm{V})$,
where the latter terms refer to the ground-state elemental phases. Of
the 64 chalcopyrites, 3/4 have negative enthalpies, suggesting many
potentially stable host materials. (We have not considered phase
stability with respect to disproportionation into compound products,
and so the terms ``stable'' and ``unstable'' should be understood here
with this caveat.)  The trend in Fig.~2 is similar, and indeed
intimately related, to that of Fig.~1: smaller lattice constants are
usually associated with higher stability. Hence, as expected, the
nitrides tend to be the most stable chalcopyrites and the antimonides
the least stable.  Note that there are important deviations from this
trend: for example, nearly all the carbides are unstable.

The solubility of Mn in a host chalcopyrite is an extremely important
issue for two reasons. The first pertains to the well-known tendency,
within Mn-doped DMS materials, toward phase separation with increasing
Mn content\cite{ohno_science_1998a,maekawa2002a}.  This tendency is
reflected in the very low equilibrium solubility limit of Mn in, for
example, III-V hosts---typically of order parts per thousand or
less. These limits can be greatly exceeded, by 2--3 orders of
magnitude, by relying on kinetic barriers to maintain the metastable
phase.  It is not feasible to address this issue theoretically for the
chalcopyrites. Instead, we propose using the theoretical solubility
limit of Mn as a rough proxy for the metastability doping limit.  The
solubility limit of Mn, $x_{\rm sol}$, is determined by its impurity
formation energy, $\Delta H_f(\textrm{Mn})$, simply according to
$x_{\rm sol}=\exp(-\Delta H_f(\textrm{Mn})/kT)$. In this formulation,
a negative formation energy implies that spontaneous incorporation of
Mn impurities will be limited only by kinetics\cite{zhang_j_phys_cond_mat_2002a}.

\begin{figure}[t]
\resizebox{86mm}{!}{\includegraphics{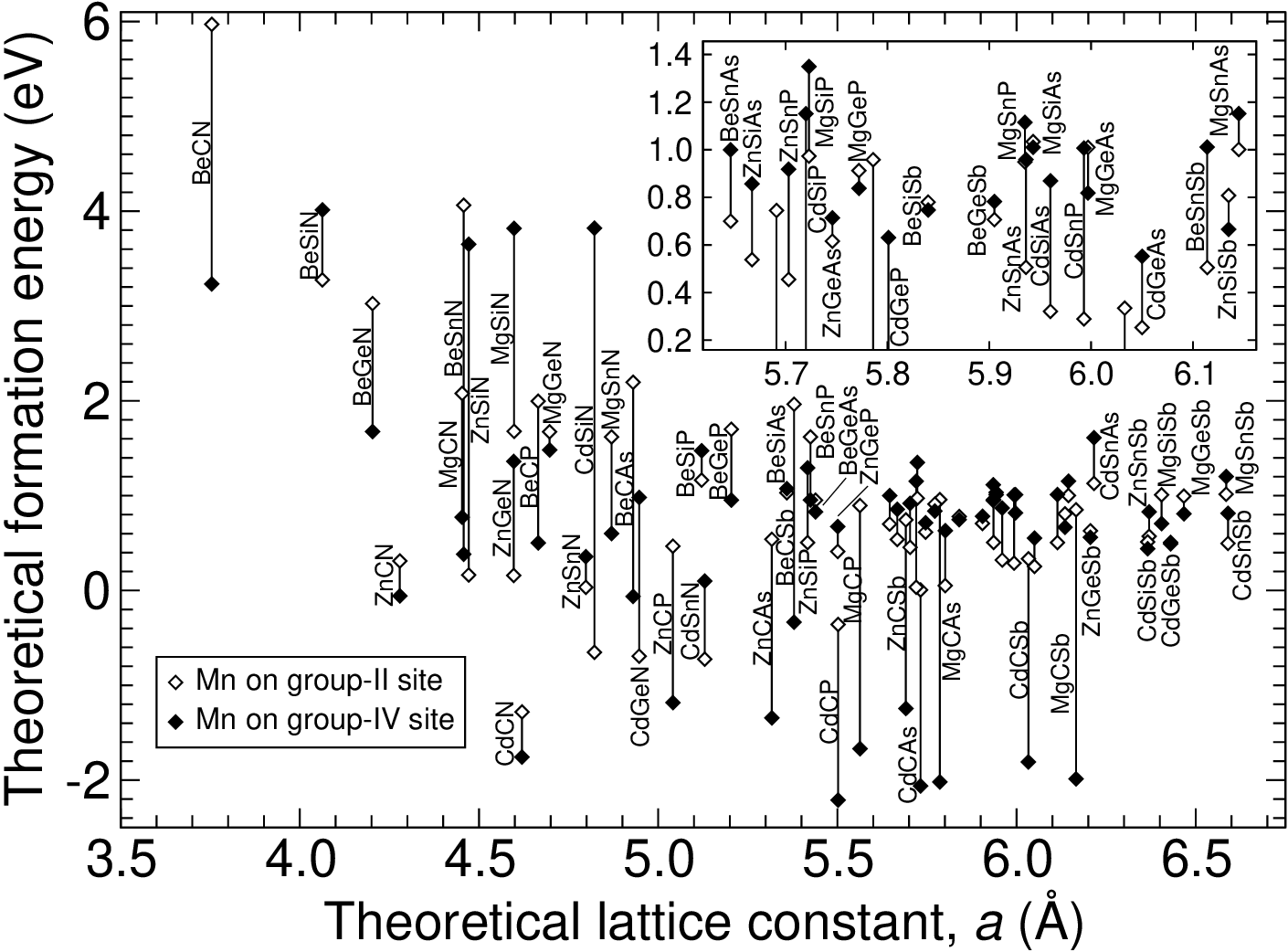}}
\caption{Theoretical impurity formation energy of substitutional Mn
vs.~lattice constant of host chalcopyrite. Chemical potentials were defined by assuming
the simultaneously II-rich and IV-rich condition.}
\end{figure}
  
\begin{figure*}
\resizebox{150mm}{!}{\includegraphics{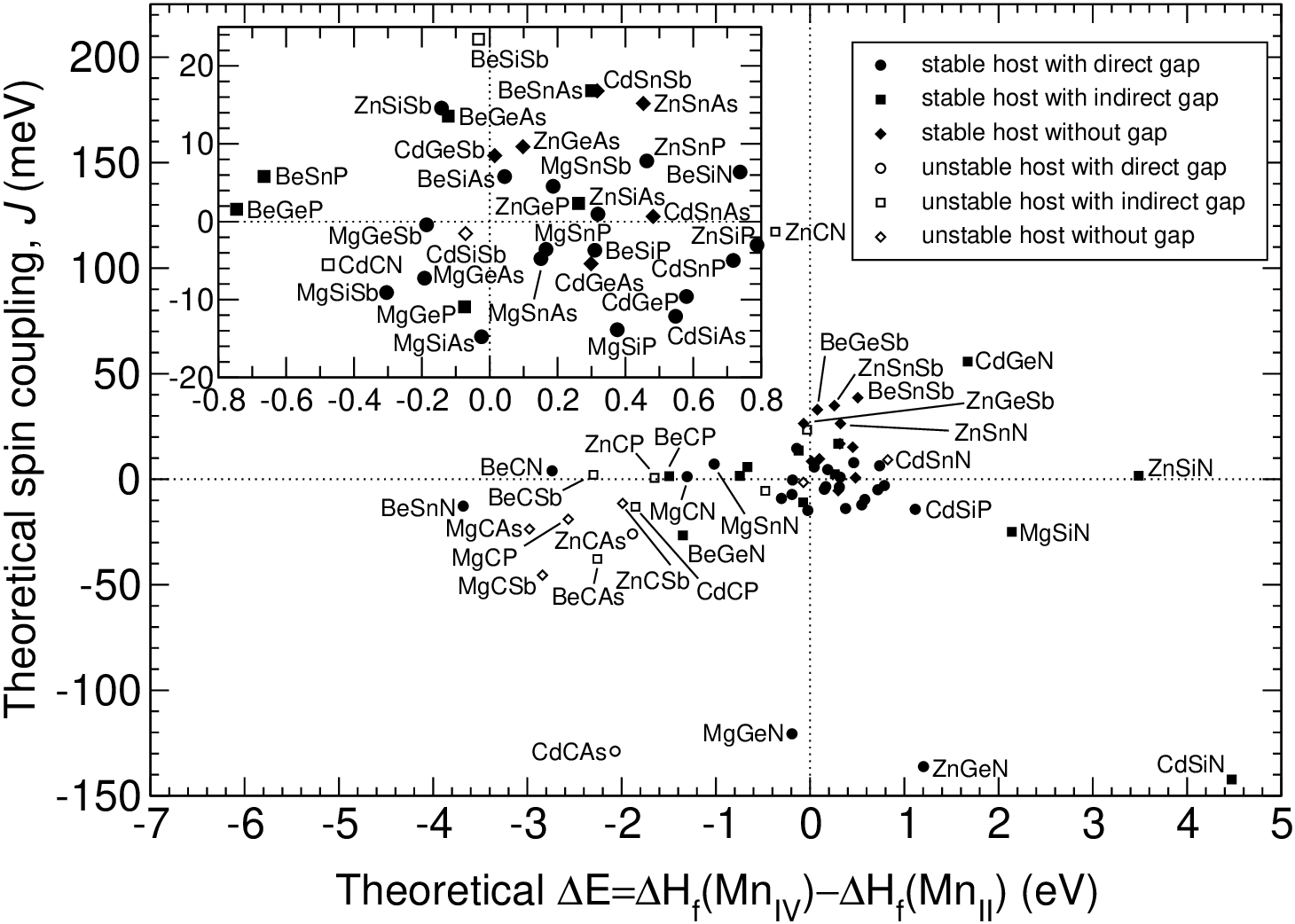}}
\caption{Theoretical spin coupling between Mn$_{\rm IV}$ 
vs.~relative impurity formation energy of Mn$_{\rm IV}$. Negative
spin coupling implies ferromagnetic ordering is favored; negative
$\Delta E$ implies substitutional Mn on the group-IV site is
favored. Chemical potentials were defined by assuming the
simultaneously II-rich and IV-rich condition. One host, CdCSb$_2$,
falls outside the plot range ($\Delta E=-2.1$ eV, $J=-350$ meV) but
is thermodynamically unstable.}
\end{figure*}

The second reason concerns ferromagnetism per se.  In general, the
solubility limits for Mn substituting on the group-II and -IV sites
will be different. Mahadevan and Zunger have shown that, in the case
of CdGeP$_2$, substitutional Mn$_{\rm II}$ leads to antiferromagnetic
interactions while Mn$_{\rm IV}$ leads to ferromagnetic
interactions\cite{mahadevan_phys_rev_lett_2002a}. Hence the II-IV
solubility difference will greatly influence the stability, or even
existence, of the ferromagnetic phase, as well as the attainable Curie
temperatures. There are two other important issues, which we do not
address here. One is the possibility of Mn occupying interstitial
sites (which is increasingly likely for hosts with small lattice
constants\cite{erwin_phys_rev_b_2003a}) or forming complexes with
other Mn or native defects; if electrically active, these can have
consequences for ferromagnetism. The other is the possibility that even
substitutional Mn may be subject to a Jahn-Teller distortion, which
can affect the magnetic interactions as discussed in
Ref.~[\onlinecite{kacman_semicond_sci_technol_2001a}].

It is important to realize that the impurity formation energy (and
hence the solubility limit) in a multicomponent host is not uniquely
defined, but depends on the growth environment---in particular, on the
energy $\mu$ required to transfer one atom from a reservoir into the
host compound\cite{vandewalle_phys_rev_b_2001a}.  For the ternary chalcopyrites there are three such
chemical potentials,
\{$\mu_{\rm II}$, $\mu_{\rm IV}$,
$\mu_{\rm V}$\}, which are related by $\mu_{\rm II}+\mu_{\rm
IV}+2\mu_{\rm V}=\Delta H_f(\textrm{II-IV-V}_2).$ The formation
energies shown in Fig.~3 were evaluated for $\mu_{\rm II}$ and
$\mu_{\rm IV}$ at their highest allowed values, corresponding to a
simultaneously II-rich and IV-rich growth environment.  This condition
can often be realized experimentally, unless a more stable compound
intervenes.  With this choice, half of
the chalcopyrite hosts (including all of the carbides) prefer Mn on
the group-II site, while half (including most of the silicides) prefer
the group-IV site.  For both the II and IV sites, there is the
expected trend of larger lattice constants leading to smaller impurity
formation energies---and hence to higher solubilities.  This trend
begins to disappear for lattice constants larger than $\sim$6 \AA; in
this regime there is enough room available for the Mn impurity that
the formation energies become essentially independent of both site and
host.

In the remainder of this work we assume that the prospects for
ferromagnetism are best served when Mn substitutes on the group-IV
site\cite{mahadevan_phys_rev_lett_2002a}. The doping can be biased in
this direction by changing the growth environment to the
simultaneously II-rich and IV-poor condition, which reduces the
relative Mn$_{\rm IV}$ formation energy, $\Delta E= \Delta
H_f(\textrm{Mn$_{\rm IV}$})-\Delta H_f(\textrm{Mn$_{\rm II}$})$, to
its minimum physically allowed value.  The magnitude of the reduction
depends on the nature of other possible intervening phases
II$_x$IV$_y$V$_z$, which we do not pursue here (see
Ref.~[\onlinecite{zhao_phys_rev_b_2004a}] for a related discussion).  In the
best-case scenario, the limiting IV-poor condition will lead to a
change in $\Delta E$ by an amount $\Delta H_f(\textrm{II-IV-V}_2)$,
relative to the values shown in Fig.~3.  Using our calculated
enthalpies, we find this limiting IV-poor condition favors Mn on the
group-IV site in 50 of the 64 hosts (the carbides are again the
dominant exception). Thus we conclude that under suitable growth
conditions, substitutional Mn can be forced to preferentially occupy
the group-IV site in the majority of chalcopyrite hosts.

Even in such favorable cases, one expects some fractional occupation
of Mn on group-II sites. Since this doping is isoelectronic it is
unlikely to produce either holes or electrons, and thus will probably
require additional doping to order ferromagnetically.
A more pressing question is whether Mn doping on the
group-IV sites necessarily leads to ferromagnetism in every
II-IV-V$_2$ chalcopyrite. And, for those hosts that can be made
ferromagnetic, it is of great interest to know what Curie temperatures
one can expect.

These are very difficult questions to address comprehensively, in a
materials-specific framework, for any random magnetic alloy. Instead,
we adopt again our earlier view and focus on a simple proxy for
ferromagnetism, namely the nearest-neighbor Heisenberg coupling, $J$,
for a specific distribution of Mn in each host. In the mean-field
theory of Heisenberg ferromagnetism, the Curie temperature is simply
proportional to $|J|$, with a proportionality constant depending on
the spatial distribution of Mn. Nevertheless, we caution that while
qualitative trends in $J$ are likely to be meaningful within this
materials class, the quantitative prediction of Curie temperatures in
ferromagnetic semiconductors remains a difficult problem.  Here we
choose a fixed, experimentally reasonable Mn concentration of 12.5\%
(see, for example, Ref.~[\onlinecite{cho_phys_rev_lett_2002a}]), and
examine how $J$ varies across different hosts. In a related study of
the I-III-VI$_2$ chalcopyrite CuGaSe$_2$, Picozzi {\it et al.}~showed
that $|J|$ increases with Mn
concentration\cite{picozzi_phys_rev_b_2002a}.  Our method for
computing $J$ is the same as in
Ref.~[\onlinecite{picozzi_phys_rev_b_2002a}], and gives nearly
identical results for CuGaSe$_2$ with 12.5\% Mn.

Surprisingly, we find that ferromagnetic alignment is favored ($J$ is
negative) in only half of the 64 chalcopyrites.  Three main factors
determine the behavior of $J$ across different hosts. The first is the
group-II element: the majority of Mg- and Cd-containing
hosts favor ferromagnetic alignment, while the majority of Be- and
Zn-containing hosts favor antiferromagnetic alignment.  The second
factor is the group-IV element: for many 
combinations of II and V elements, we find the ordering $J({\rm
C})<J({\rm Si})<J({\rm Ge})<J({\rm Sn})$, except in the nitrides
(see Supplementary Information, Fig.~S1).  The group-V
element plays the third---and lesser---role, with the
ordering $J({\rm P})<J({\rm As})<J({\rm Sb})$ obtained for most cases
except the carbides (see Supplementary Information, Fig.~S2).  None of
these trends can be attributed to the approximate scaling, $J\sim
a^{-3}$, predicted by the mean-field solution of the Zener
model\cite{dietl_phys_rev_b_2001a}; indeed, across this class of
chalcopyrites we find no universal dependence of $J$ on lattice
constant (see Supplementary Information, Fig.~S3).  Nor do we find any
simple correlation between $J$ and the bandgap; such a correlation
might have been expected if the small-gap hosts reflected Zener
physics (with modest Curie temperatures) and the large-gap hosts
double-exchange physics (with higher Curie
temperatures)\cite{pearton_j_appl_phys_2003a}.  

Fig.~4 shows the computed values of $J$ plotted against the relative
Mn$_{\rm IV}$ impurity formation energy, $\Delta E$. We do not imply
any correlation between these two quantities, but rather display them
together because each is centrally important for designing new
ferromagnetic chalcopyrites. In particular, the lower-left quadrant
contains chalcopyrites for which occupation by, and ferromagnetic
alignment of, Mn on the group-IV site is favorable under
experimentally plausible II- and IV-rich growth conditions.  This
conservatively defined sweet spot contains just 19 chalcopyrites out
of the full set of 64. Of these, only eight are thermodynamically
stable: four are lattice-matched to a mainstream semiconductor
(BeSnN$_2$, MgGeP$_2$, MgSiAs$_2$, MgGeAs$_2$) and four are not
(BeGeN$_2$, MgSiSb$_2$, MgGeN$_2$, MgGeSb$_2$). Seven of these eight
have values of $J$ comparable to, or larger than, that of CdGeP$_2$
and thus can be expected to have similar Curie temperatures---of order
room temperature or larger.  This small set of materials should
provide a propitious starting point for more detailed experimental and
theoretical scrutiny.

One can also consider the best-case scenario described earlier and
consider $\Delta E$ in the limiting IV-poor growth condition. This
leads to horizontal shifts, by $\Delta H_f$, of the points in
Fig.~4---leftward for negative enthalpies and rightward for
positive---and thereby moves 13 more chalcopyrites into the lower left
sweet spot.  Ten of these are lattice-matched to mainstream semiconductors
(MgSiN$_2$, MgSiP$_2$, MgSnP$_2$, MgSnAs$_2$, ZnSiP$_2$, ZnGeN$_2$,
CdSiP$_2$, CdSiAs$_2$, CdGeAs$_2$, CdSnP$_2$), and three are not
(BeSiP$_2$, CdSiN$_2$, CdGeP$_2$); all are thermodynamically
stable.

Experimentally, CdGeP$_2$ and ZnSnAs$_2$ have ferromagnetic ground
states with comparable Curie temperatures, while ZnGeP$_2$ has an
antiferromagnetic ground state (ferromagnetism disappears below 47 K
[\onlinecite{cho_phys_rev_lett_2002a}]).  Our results, which presume
Mn doping on the group-IV site, predict CdGeP$_2$ to be ferromagnetic
but ZnSnAs$_2$ and ZnGeP$_2$ to be antiferromagnetic. Reconciling these
findings is problematic, because the experimental group-II
and -IV chemical potentials---and thereby the favored Mn site---are in
general not known.  The available evidence points to Mn$_{\rm IV}$
substitution in CdGeP$_2$
[\onlinecite{mahadevan_phys_rev_lett_2002a}], Mn$_{\rm II}$ in
ZnGeP$_2$ [\onlinecite{cho_phys_rev_lett_2002a}], and mixed occupation
in ZnSnAs$_2$ [\onlinecite{choi_solid_state_comm_2002a}]. Future work
addressing the magnetic interactions between Mn on group-II sites is
in progress.

Finally, we address the prospects for using ferromagnetic
chalcopyrites as sources of spin-polarized current for injection into
nonmagnetic semiconductors. Many issues play a role in this
process, including the spin polarization of states at the
Fermi level, the band velocities of those states, 
native defects and other structural inhomogeneities, and
spin scattering at the interface\cite{zutic_rev_mod_phys_2004a}.  Here
we focus on the first of these. For dilute Mn doping (12.5\% and
6.25\%, and presumably below this as well), all of the 21 stable
ferromagnetic chalcopyrites identified above are ``idealized
half-metals,'' by which we mean at that at zero temperature and for
an ordered arrangement of Mn, there are states at the Fermi level only
in one spin channel. The effects of both finite temperature
and site disorder will generally reduce the spin polarization from
this idealized case.  At higher Mn concentration (50\%), 
half-metallicity persists for all but two of these 21 hosts
(BeSnN$_2$, CdSiN$_2$).  Even when Mn completely occupies the group-IV
sublattice, the resulting stoichiometric compounds are half-metallic
in all cases except for the nitrides. Thus the prospects for using
ferromagnetic chalcopyrites as spin-polarized sources appear 
quite favorable.

In summary, we have examined theoretically the prospects for
ferromagnetism within the class of all possible II-IV-V$_2$
chalcopyrites having constituents chosen from the first four rows of
the Periodic Table. As expected for semiconductors, we find that the
electronic properties of the host materials (band gap and enthalpy of
formation) are closely related to their structural properties (lattice
constant, $a$).  Contrary to conventional wisdom, we do not find that
Mn is generally more soluble on the group-II site than on the group-IV
site; instead, the relative site solubility depends on both the nature
of the host (e.g.~carbide vs.~silicide) and the choice of growth
condition, while the absolute Mn solubility is largely determined by
the host lattice constant. Our results for the Heisenberg coupling
between Mn show that the Curie temperatures expected in Mn-doped
chalcopyrites do not follow the scaling with lattice constant
predicted by the Zener model; nor do they show any systematic
variation with band gap, as the double-exchange model would predict.
By identifying those chalcopyrites that simultaneously exhibit thermodynamic
stability, favorable Mn doping site, and ferromagnetic Mn interactions, we
identify two small sets of chalcopyrites that show excellent prospects
for stable ferromagnetism under realistic and attainable experimental
conditions.

This work was supported by ONR and the DARPA SpinS program.
I.\v{Z}. acknowledges financial support from the National Research Council.
Discussions with G.A.~Medvedkin are gratefully acknowledged.
Computations were performed at the DoD Major Shared Resource Center at
ASC. 

{\bf Methods.} {\small All the numerical results reported here are based on
density-functional theory in the generalized-gradient
approximation\cite{perdew_phys_rev_b_1992a}, using ultrasoft
pseudopotentials as implemented in {\sc vasp}
\cite{kresse_phys_rev_b_1993a,kresse_phys_rev_b_1996a}.
For each host chalcopyrite, the plane-wave cutoff was separately
determined by the three constituent elements (four if Mn was
included).  For the undoped hosts, equilibrium lattice parameters and
the internal coordinate were optimized using a 4$\times$4$\times$4
Monkhorst-Pack sampling of the Brillouin zone.  Band gaps were
evaluated by identifying band edges based on a 12$\times$12$\times$12
sampling that included the zone center and high-symmetry points.

For enthalpies of formation, the total energies of the elemental
phases were calculated using the following structures, which are
either ground-state phases or energetically very close thereto:
hcp (Be, Mg, Cd, Zn); diamond (C, Si, Ge); white-tin (Sn); molecular dimer
(N); black phosphorous (P); $\alpha$As (As, Sb). 

Mn impurity formation energies were computed using simple cubic (or,
for non-ideal $c/a$, tetragonal) supercells of 64 host atoms.  To
properly represent the dilute impurity limit, all atomic positions
were relaxed with fixed ideal supercell lattice parameters, within the
constraint that Mn remain on-center. A single Monkhorst-Pack $k$-point
was used. Chemical potentials were defined with respect to their
thermodynamic upper limit determined by the elemental phases listed
above.  For the Mn chemical potential, a non-magnetic fcc structure
was used to approximate the more complicated $\alpha$Mn ground-state
phase.

Heisenberg spin couplings for 12.5\% Mn-doped chalcopyrites were
computed using the method described in
Ref.~\onlinecite{picozzi_phys_rev_b_2002a}.  The lattice parameters
and internal coordinate of each supercell were assumed to depend
linearly on the Mn content (Vegard's law), and were interpolated using
lattice parameters and internal coordinate calculated for the fully
Mn$_{\rm IV}$-substituted host.  Two Mn per 64-atom supercell were
then arranged on an uniform lattice and full atomic relaxation
was performed.  Total energies for two spin configurations were
computed---ferromagnetic, and antiferromagnetic with $[$100$]$
ordering wavevector---whereupon the energy difference gives $J$ within
the nearest-neighbor Heisenberg model.}


\begin{thebibliography}{10}

\bibitem{ohno_science_1998a}
Ohno, H.
\newblock Making nonmagnetic semiconductors ferromagnetic.
\newblock {\em Science}{ \bf 281}, 951--956 (1998).

\bibitem{maekawa2002a}
Maekawa, S. and {Shinjo, T. (Eds.)}.
\newblock {\em Spin Dependent Transport in Magnetic Nanostructures}.
\newblock Taylor and Francis, New York,  (2002).

\bibitem{zutic_rev_mod_phys_2004a}
\v{Z}uti\'c, I., Fabian, J., and {Das Sarma}, S.
\newblock Spintronics: Fundamentals and applications.
\newblock {\em Rev. Mod. Phys.}{ \bf 76} (2004).
\newblock (in press).

\bibitem{dietl_nature_mater_2003a}
Dietl, T.
\newblock Functional ferromagnetics.
\newblock {\em Nature Mater.}{ \bf 2}, 646--648 (2003).

\bibitem{dietl_science_2000a}
Dietl, T., Ohno, H., Matsukura, F., Cibert, J., and Ferrand, D.
\newblock Zener model description of ferromagnetism in zinc-blende magnetic
  semiconductors.
\newblock {\em Science}{ \bf 287}, 1019 (2000).

\bibitem{konig_phys_rev_lett_2000a}
K{\"o}nig, J., Lin, H.-H., and MacDonald, A.~H.
\newblock Theory of diluted magnetic semiconductor ferromagnetism.
\newblock {\em Phys. Rev. Lett.}{ \bf 84}, 5628--5631 (2000).

\bibitem{chattopadhyay_phys_rev_lett_2001a}
Chattopadhyay, A., {Das Sarma}, S., and Millis, A.~J.
\newblock Transition temperature of ferromagnetic semiconductors: a dynamical
  mean field study.
\newblock {\em Phys. Rev. Lett.}{ \bf 87}, 227202 (2001).

\bibitem{nagaev2002a}
Nagaev, E.~L.
\newblock {\em {Colossal Magnetoresistance and Phase Separation in Magnetic
  Semiconductors}}.
\newblock Imperial College Press, London,  (2002).

\bibitem{medvedkin_japanese_journal_of_applied_physics_2000a}
Medvedkin, G.~A., Ishibashi, T., Hayata, T.~N., Hasegawa, Y., and Sato, K.
\newblock Room temperature ferromagnetism in novel diluted magnetic
  semiconductor {Cd$_{1-x}$Mn$_{x}$GeP$_{2}$}.
\newblock {\em Jpn. J. Appl. Phys., Part 2}{ \bf 39}, L949--951 (2000).

\bibitem{medvedkin_journal_of_crystal_growth_2002a}
Medvedkin, G.~A., Hirose, K., Ishibashi, T., Nishi, T., Voevodin, V.~G., and
  Sato, K.
\newblock New magnetic materials in {ZnGeP$_2$Mn} chalcopyrite system.
\newblock {\em Journal of Crystal Growth}{ \bf 236}, 609--612 (2002).

\bibitem{cho_phys_rev_lett_2002a}
Cho, S., Choi, S., Cha, G.-B., Hong, S.~C., Kim, Y., Zhao, Y.-J., Freeman,
  A.~J., Ketterson, J.~B., Kim, B.~J., and Kim, Y.~C.
\newblock Room-temperature ferromagnetism in {Zn$_{1-x}$Mn$_x$GeP$_2$}
  semiconductors.
\newblock {\em Phys. Rev. Lett.}{ \bf 88}, 257203 (2002).

\bibitem{choi_solid_state_comm_2002a}
Choi, S., Cha, G.-B., Hong, S.~C., Cho, S., Kim, Y., Ketterson, J.~B., Jeong,
  S.-Y., and Yi, G.~C.
\newblock Room-temperature ferromagnetism in chalcopyrite {Mn}-doped
  {ZnSnAs$_2$} single crystals.
\newblock {\em Solid State Comm.}{ \bf 122}, 165--167 (2002).

\bibitem{perdew_phys_rev_b_1992a}
Perdew, J.~P., Chevary, J.~A., Vosko, S.~H., Jackson, K.~A., Pederson, M.~R.,
  Singh, D.~J., and Fiolhais, C.
\newblock Atoms, molecules, solids, and surfaces: Applications of the
  generalized gradient approximation for exchange and correlation.
\newblock {\em Phys. Rev. B}{ \bf 46}, 6671--6687 (1992).

\bibitem{kresse_phys_rev_b_1993a}
Kresse, G. and Hafner, J.
\newblock Ab initio molecular dynamics for liquid metals.
\newblock {\em Phys. Rev. B}{ \bf 47}, 558--561 (1993).

\bibitem{kresse_phys_rev_b_1996a}
Kresse, G. and Furthm{\"u}ller, J.
\newblock Efficient iterative schemes for ab initio total-energy calculations
  using a plane-wave basis set.
\newblock {\em Phys. Rev. B}{ \bf 54}, 11169--11186 (1996).

\bibitem{mahadevan_phys_rev_lett_2002a}
Mahadevan, P. and Zunger, A.
\newblock Room-temperature ferromagnetism in {Mn}-doped semiconducting
  {CdGeP$_2$}.
\newblock {\em Phys. Rev. Lett.}{ \bf 88}, 047205 (2002).

\bibitem{continenza_phys_rev_b_1992a}
Continenza, A., Massidda, S., Freeman, A.~J., de~Pascale, T.~M., Meloni, F.,
  and Serra, M.
\newblock Structural and electronic properties of narrow-gap {ABC$_2$}
  chalcopyrite semiconductors.
\newblock {\em Phys. Rev. B}{ \bf 46}, 10070--10077 (1992).

\bibitem{zhang_j_phys_cond_mat_2002a}
Zhang, S.~B.
\newblock The microscopic origin of the doping limits in semiconductors and
  wide-gap materials and recent developments in overcoming these limits: a
  review.
\newblock {\em J.~Phys.:~Condens.~Matter}{ \bf 14}, R881--R903 (2002).

\bibitem{erwin_phys_rev_b_2003a}
Erwin, S.~C. and Hellberg, C.~S.
\newblock Predicted absence of ferromagnetism in manganese-doped diamond.
\newblock {\em Phys. Rev. B}{ \bf 68}, 245206 (2003).

\bibitem{kacman_semicond_sci_technol_2001a}
Kacman, P.
\newblock Spin interactions in diluted magnetic semiconductors and magnetic
  semiconductor structures.
\newblock {\em Semicond.~Sci.~Technol.}{ \bf 16}, R25--R39 (2001).

\bibitem{vandewalle_phys_rev_b_2001a}
{Van de Walle}, C.~G., Limpijumnong, S., and Neugebauer, J.
\newblock First-principles studies of beryllium doping of {GaN}.
\newblock {\em Phys. Rev. B}{ \bf 63}, 245205 (2001).

\bibitem{zhao_phys_rev_b_2004a}
Zhao, Y.-J. and Zunger, A.
\newblock Site preference for {Mn} substitution in spintronic {CuM$^{\rm
  III}$X$_2^{\rm VI}$} chalcopyrite semiconductors.
\newblock {\em Phys. Rev. B}{ \bf 69}, 075208 (2004).

\bibitem{picozzi_phys_rev_b_2002a}
Picozzi, S., Zhao, Y.-J., Freeman, A.~J., and Delley, B.
\newblock Mn-doped {CuGaS$_2$} chalcopyrites: an ab initio study of
  ferromagnetic semiconductors.
\newblock {\em Phys. Rev. B}{ \bf 66}, 205206 (2002).

\bibitem{dietl_phys_rev_b_2001a}
Dietl, T., Ohno, H., and Matsukura, F.
\newblock Hole-mediated ferromagnetism in tetrahedrally coordinated
  semiconductors.
\newblock {\em Phys. Rev. B}{ \bf 63}, 195205 (2001).

\bibitem{pearton_j_appl_phys_2003a}
Pearton, S.~J., Abernathy, C.~R., Overberg, M.~E., Thaler, G.~T., Norton,
  D.~P., Theodoropoulou, N., Hebard, A.~F., Park, Y.~D., Ren, F., Kim, J., and
  Boatner, L.~A.
\newblock Wide band gap ferromagnetic semiconductors and oxides.
\newblock {\em J. Appl. Phys.}{ \bf 93}, 1--13 (2003).

\end{thebibliography}

\newpage

\onecolumngrid
{\bf Supplementary Information}

\begin{figure*}[h]
\resizebox{120mm}{!}{\includegraphics{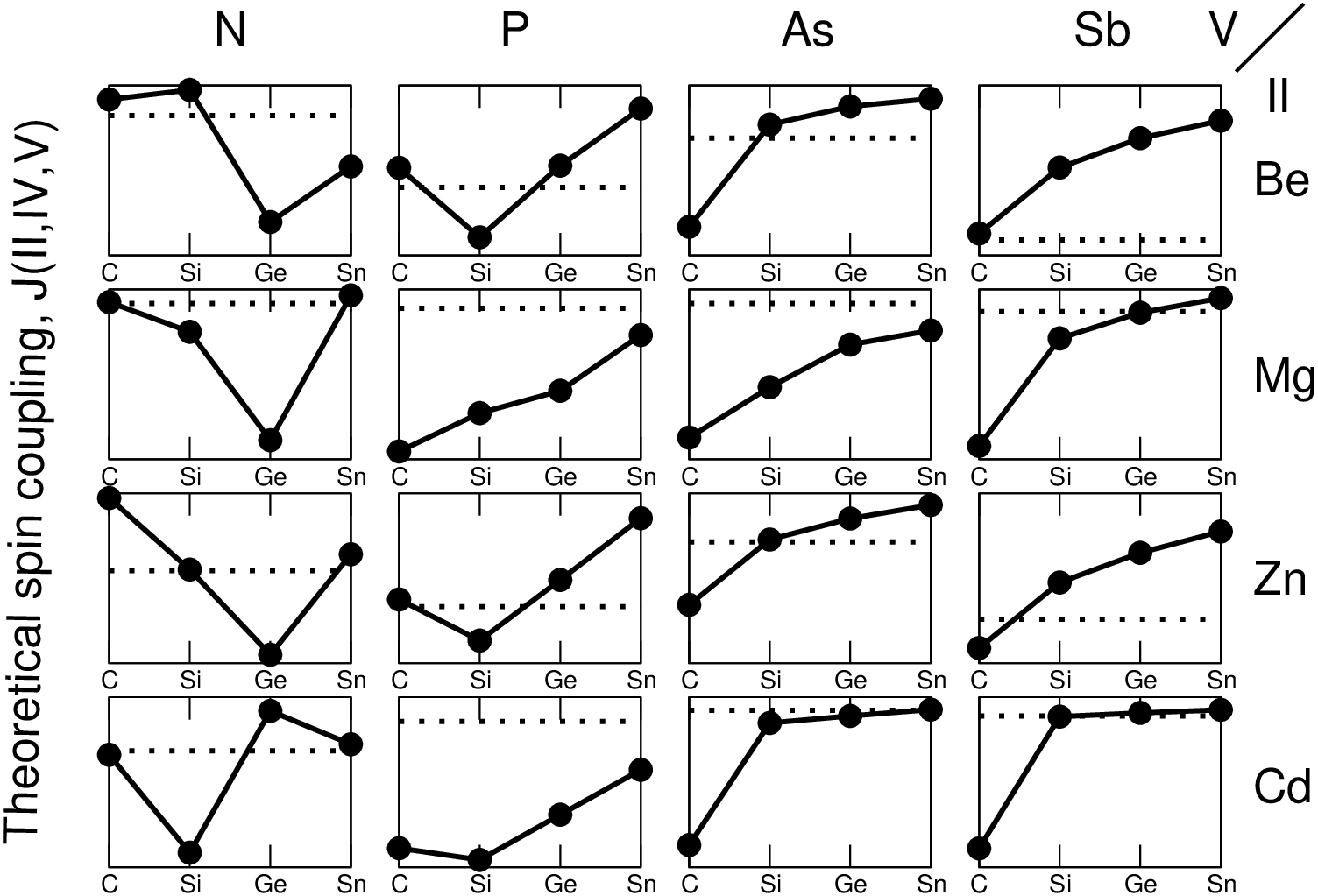}}
\caption{[FIG. S1] Theoretical Mn$_{\rm IV}$ spin coupling
vs.~group-IV element, for each of 16 sets of II-IV-V$_2$
chalcopyrites.  Results are plotted with arbitrary units to emphasize
trends; the dotted lines denote $J=0$.  For many combinations of
II and V elements, the ordering $J({\rm C})<J({\rm Si})<J({\rm
Ge})<J({\rm Sn})$ is found, except in the nitrides.}
\end{figure*}

\begin{figure*}[h]
\resizebox{120mm}{!}{\includegraphics{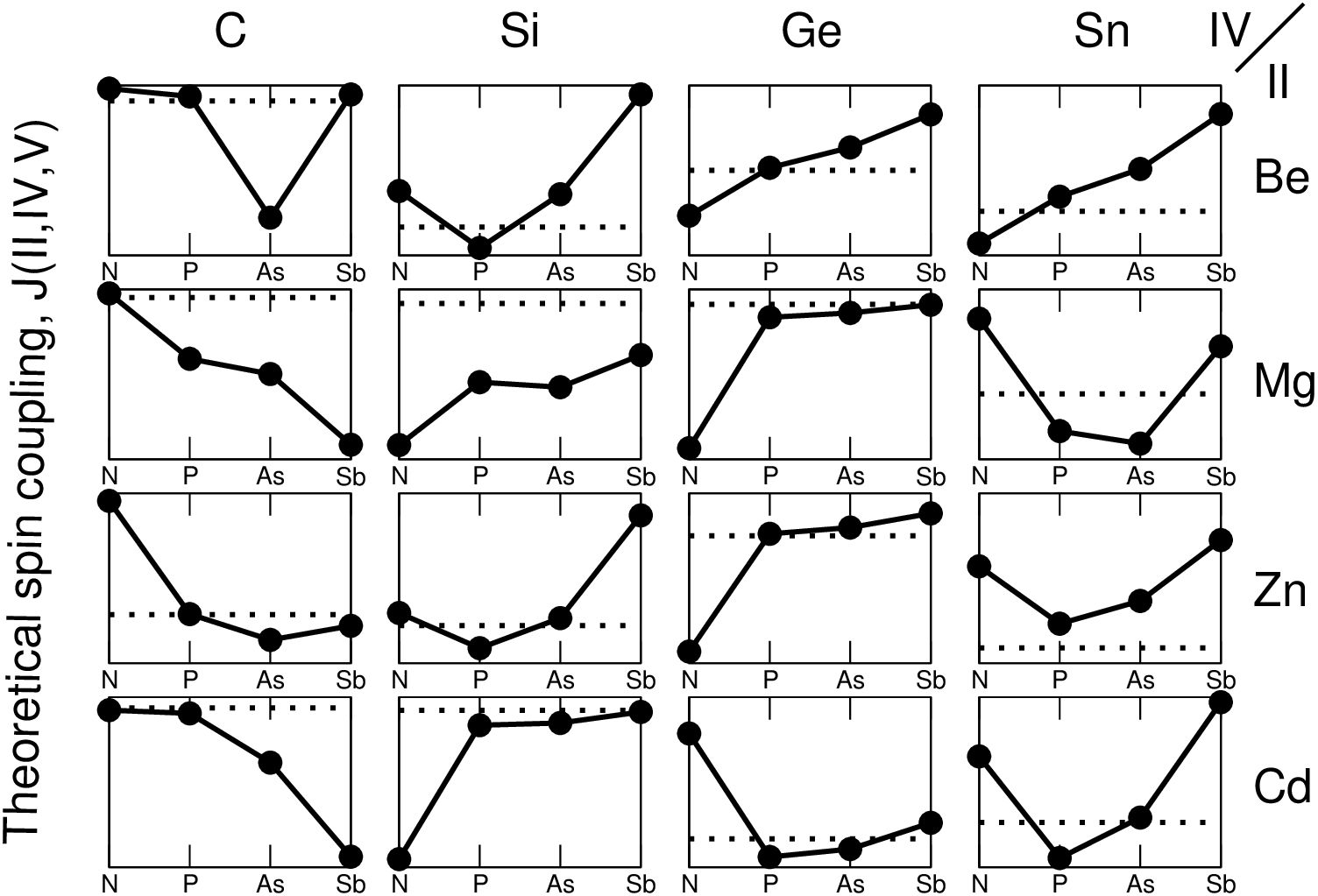}}
\caption{[FIG. S2] Theoretical Mn$_{\rm IV}$ spin coupling
vs.~group-V element, for each of 16 sets of II-IV-V$_2$
chalcopyrites. 
 Results are plotted with arbitrary units to emphasize trends; the dotted
lines denote $J=0$.  The ordering $J({\rm P})<J({\rm As})<J({\rm Sb})$
is obtained for most cases, except in the carbides.}
\end{figure*}

\begin{figure*}[h]
\resizebox{120mm}{!}{\includegraphics{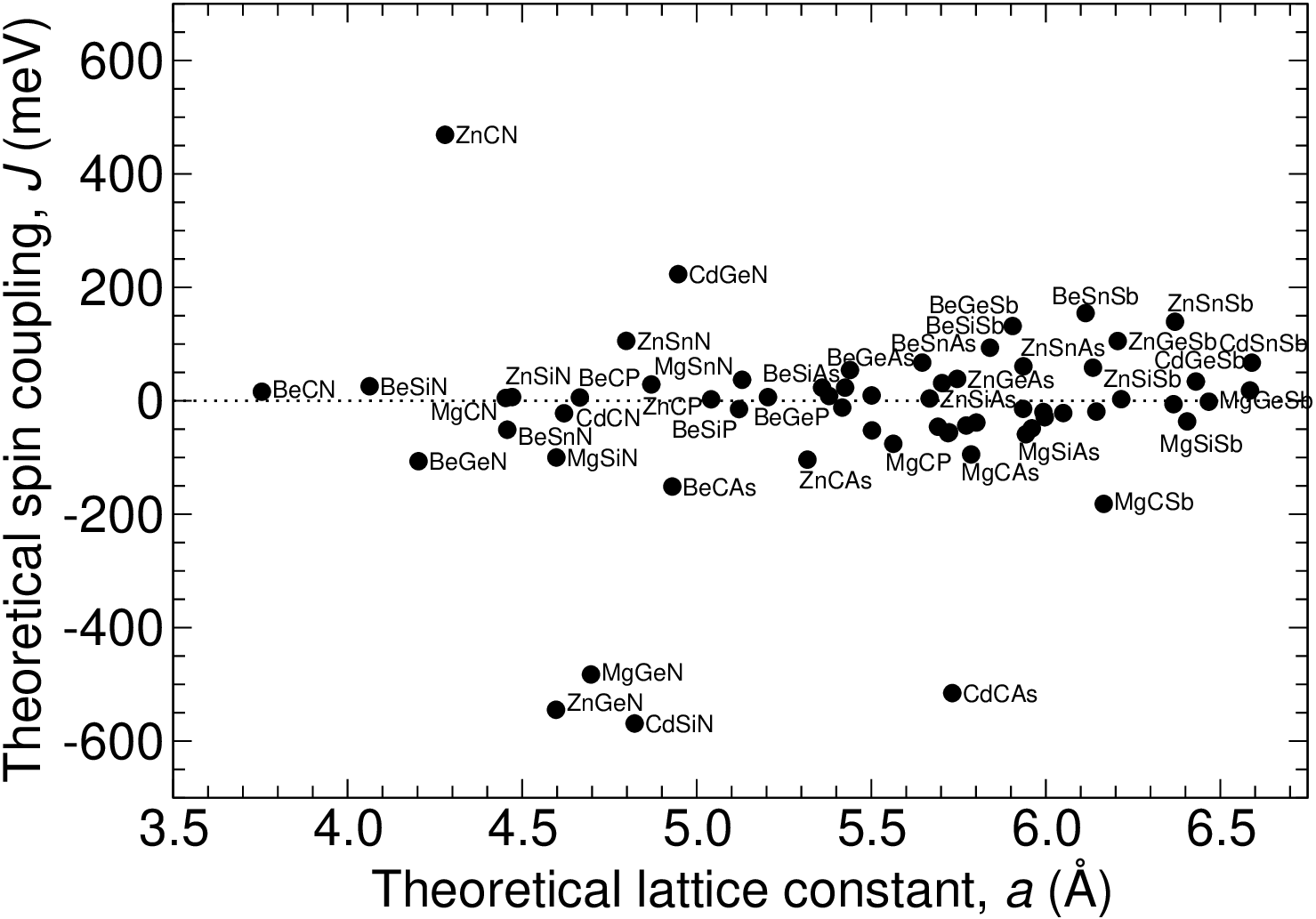}}
\caption{[FIG. S3] Theoretical Mn$_{\rm IV}$ spin coupling
vs.~lattice constant of host chalcopyrite. No evidence is observed for
the approximate scaling, $J\sim a^{-3}$, predicted by the mean-field
solution of the Zener model.}
\end{figure*}

\end{document}